\documentclass[proceedings, preprint]{rmaa}

\usepackage{paralist}
\usepackage{txfonts}
\usepackage{natbib}
\usepackage{psfrag,color}

\SetYear{2007}
\SetConfTitle{XII Latin-American Regional meeting of IAU}

\title{The vertical velocity dispersion profile of the Galactic thick disk} 

\author{
C.~Moni Bidin\altaffilmark{1,2},
T.~M.~Girard\altaffilmark{2},
G.~Carraro\altaffilmark{3},
R.~A.~M\'{e}ndez\altaffilmark{1},
W.~F.~van Altena\altaffilmark{2},
V.~I.~Korchagin\altaffilmark{2},
and D.~I.~Casetti-Dinescu\altaffilmark{2}
}

\altaffiltext{1}{Departamento de Astronom\'{i}a, Universidad de Chile,
Casilla 36-D, Santiago, Chile (mbidin@das.uchile.cl).}
\altaffiltext{2}{Yale University, Dept. of Astronomy,
P.O. Box 208101, New Haven, CT 06520-8101.}
\altaffiltext{3}{European Southern Observatory, 3107
Alonso de Cordova, Santiago, Chile.}

\shortauthor{Moni Bidin et al.}

\shorttitle{The vertical velocity dispersion profile of the Galactic thick disk}

\listofauthors{C. Moni Bidin, T. M. Girard, G. Carraro, R. A. M\'{e}ndez,
W. F. van Altena, V. I. Korchagin, \& D. I. Casetti-Dinescu}
\indexauthor{Moni Bidin, C.}
\indexauthor{Girard, T. M.}
\indexauthor{Carraro, G.}
\indexauthor{M\'{e}ndez, R. A.}
\indexauthor{van Altena, W. F.}
\indexauthor{Korchagin, V. I.}
\indexauthor{Casetti-Dinescu, D. I.}

\abstract{We present the results of radial velocity measurements
of 770 thick disk red giants toward the South Galactic Pole,
vertically distributed from 0.5 kpc to 5 kpc with respect to the
Galactic plane. We find a small gradient in the vertical velocity
dispersion ($\sigma_W$) of 3.8$\pm$0.8 km s$^{-1}$ kpc$^{-1}$.  Even
more noteworthy, our values of $\sigma_W$ are small compared to
literature values: in the middle of the vertical height range we find
$\sigma_{W,\mathrm{z=2kpc}}$=30 km s$^{-1}$. We found no possible
explanation for this small value of $\sigma_{W}$ in
terms of sample contamination by thin disk stars, nor by wrong
assumptions regarding the metallicity distribution and the derived
distances.}

\resumen{Presentamos resultados de mediciones de velocidad radial
para 770 estrellas gigantes rojas del disco grueso hacia el Polo
Gal\'{a}ctico Sur, distribu\'{\i}das verticalmente desde 0.5 kpc hasta
5 kpc de distancia con respecto al plano Gal\'{a}ctico. Encontramos un
peque\~no gradiente en la dispersi\'{o}n de velocidad vertical
($\sigma_W$) de 3.8$\pm$0.8 km s$^{-1}$ kpc$^{-1}$. Aun mas notable,
nuestros valores de $\sigma_W$ son peque\~{n}os comparados con valores
de literatura: en el medio del rango de distancia del plano
encontramos $\sigma_{W,\mathrm{z=2kpc}}$=30 km s$^{-1}$.  No es
posible explicar este peque\~no valor de $\sigma_{W}$ en t\'erminos de
contaminaci\'{o}n de la muestra por estrellas del disco delgado, y
tampoco a trav\'{e}s de una errada distribuci\'{o}n de metalicidad y
las distancias deducidas.}

\addkeyword{Galaxy: disk}
\addkeyword{Galaxy: fundamental parameters}
\addkeyword{Galaxy: kinematics and dynamics}

\begin{document}
\maketitle

\section{Introduction}
\label{intro}

We are undertaking a spectroscopic study of nearly 1,200 thick disk
red giant stars toward the South Galactic Pole, to analyze the
chemical and kinematical vertical structure of the thick disk
\citep{Carraro05}.

The sample is vertically distributed with respect to the Galactic
plane, probing the Galactic thick disk with unprecedented detail
from 0.5 up to 5 kpc from the Galactic plane.
Details regarding target selection in the $K$ {\it vs.} $(J$-$K)$
plane using 2MASS photometry and the distance estimation procedure
were presented by
\citet{Girard06}, who studied the
proper motions of the sample from the SPM3 catalog \citep{Girard04}.

We collected high-resolution Echelle spectra for 770 stars (two
thirds of the sample) during two observing seasons in 2005 and 2006 at
various instruments.  Radial velocities (RVs) were measured through a
cross-correlation technique \citep{Tonry79}, using as templates three
red giant RV standard stars observed in all runs. Using synthetic
templates and twilight solar spectra acquired each night we corrected
all RVs from systematic errors due to a number of factors, as for
example stars not perfectly centered in the slit/fiber and RV
variations of the templates. Errors in the measurements are in the
range 0.4--1.0 km s$^{-1}$, they were evaluated quadratically summing
up the contribution of all relevant sources (for more details, see
Moni Bidin 2008, PhD Thesis, in preparation).  The comparison with
literature RVs for 162 stars with published values reveals an
excellent agreement with no systematic trend, and the RV dispersion
measured in this subsample is the same as that deduced from the
literature RVs.

\begin{figure*}[!t]
\begin{center}
\includegraphics[width=14cm]{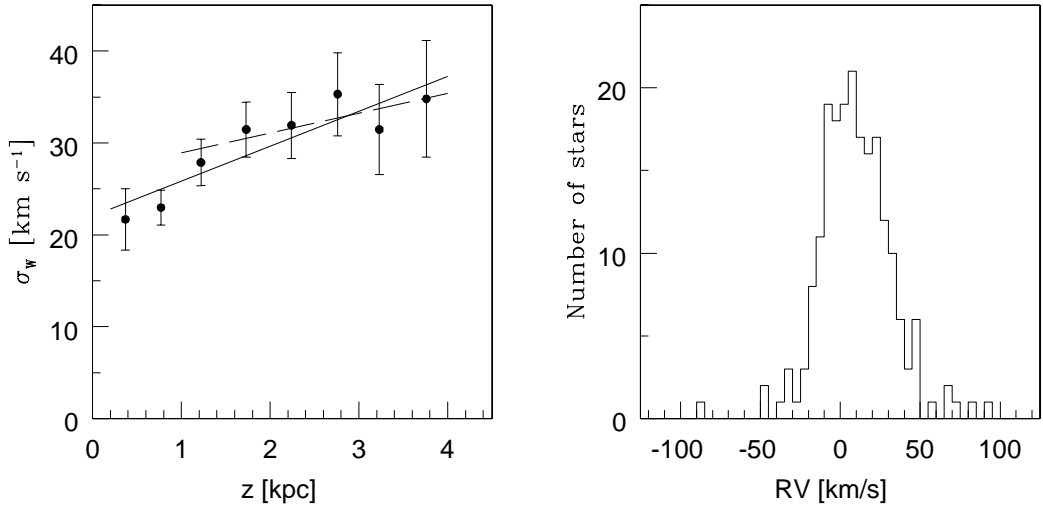}
\end{center}
\caption{{\it Left panel}:
Vertical velocity dispersion profile of the sample: The dispersion in
W-velocity is plotted as a function of distance from the
Galactic plane $z$.  The solid line indicates the least square
fit, the dashed line is the fit obtained exluding the first two data
points (suspected to be contaminated by thin disk stars).
{\it Right panel}: RV distribution of the 200 stars in the range $z$=1.5-2.5 kpc.
}
\label{plot}
\end{figure*}

\section{Results}
\label{results}

Our results are shown in Figure~\ref{plot}, where the vertical velocity
dispersion ($\sigma_W$) is plotted as a function of
distance $z$ from the Galactic plane.  In this plot we conservatively
excluded all stars with $\vert$W$\vert\geq$120 km s$^{-1}$ to avoid
ant significant residual contamination of halo stars. Our
values are systematically lower when compared to those in the
literature. Various values of $\sigma_W$ for the thick disk have been
proposed in the last two decades, varying in the range 40--70 km s$^{-1}$
\citep[see for example Table~1 in][]{Casertano90}, although recent
determinations tend to prefer values in the lower edge of this
interval (see for example \citealt{Soubiran03}, and the value adopted
by
\citealt{Bensby03}).  In contrast, we find a $\sigma_W$ always smaller
than 40 km s$^{-1}$ at all Galactic heights, with
$\sigma_{W,\mathrm{z=2kpc}}$=30 km s$^{-1}$. In the presence of a
vertical gradient (see below), the local extrapolated value would be
even lower.  We note that no effort was made to identify and remove
binary systems from the sample, a correction that would further reduce
the velocity dispersion.

We find a small vertical gradient of 3.8$\pm$0.8 km s$^{-1}$
kpc$^{-1}$, but it is mainly due to the first two bins, corresponding
to the nearest points, where a residual contamination by thin disk red
giants can not be excluded. In fact (see Figure 1), after excluding
these two first distance bins from the fit, the resulting gradient is
much shallower, and within error bars the data are consistent with a
flat profile.

We find no plausible source of thin disk contamination that can
account for our small values. Thin disk red giants contaminate (if
any) only the nearest bins. Dwarfs were efficiently excluded by a cut
in magnitude, a (conservative) cut in proper motions
\citep{Girard06} and a further inspection of all the stellar spectra.  
Moreover, \citet{Girard06} showed that in the range $z$=1--4 kpc the
density profile of the sample is well described by a single
exponential with a scale height of 783~pc, thus demonstrating that it
is dominated by thick disk stars. We found that different assumptions
on the metallicity distribution, which can lead to a wrong distance
estimate, hardly change the results: Assuming extreme fixed values for
the whole sample, we find $\sigma_{W,\mathrm{z=2kpc}}$=31 km s$^{-1}$
for [Fe/H]=$-$0.5 and $\sigma_{W,\mathrm{z=2kpc}}$=27 km s$^{-1}$ for
[Fe/H]=$-$1.1.

It is worth noting that our sample selection include stars in a rather
wide metalicity interval, but higher metallicities are preferred
\citep[see the isocrones in Figure~1 of][]{Girard06}. Hence, we conclude that
either the thick disk is, as a whole, a stellar population
kinematically cooler than believed so far, or that there exists some
$\sigma_W$-metallicity relation, in the sense of the metal-rich
component being kinematically cooler.  This conclusion was already
proposed by \citet{Schuster06}, who argued for the existence of two
thick disk components, although their derived $\sigma_W$ for the
kinematically cooler population is still high with respect to ours.
Moreover, it is becoming evident that bona-fide thick disk stars
extend at least up to solar metallicities \citep{Bensby07}.

\vspace{0.3cm}
{\small
{\it
CMB attendance to the meeting was funded by Conicyt and by the LOC.
RAM acknowledge support by the Chilean Centro de
Astrof\'{\i}sica FONDAP (No. 15010003).
}}

\end{document}